# On Robot Revolution and Taxation

Tshilidzi Marwala

University of Johannesburg

South Africa

Email: tmarwala@uj.ac.za

**Abstract**

Advances in artificial intelligence are resulting in the rapid automation of the work force. The tools that are used to automate are called robots. Bill Gates proposed that in order to deal with the problem of the loss of jobs and reduction of the tax revenue we ought to tax the robots. The problem with taxing the robots is that it is not easy to know what a robot is. This article studies the definition of a robot and the implication of advances in robotics on taxation. It is evident from this article that it is a difficult task to establish what a robot is and what is not a robot. It concludes that taxing robots is the same as increasing corporate tax.

## 1. Introduction

The first industrial revolution gave us the steam engine, the second electricity as well as the assembly line and the third electronics. The fourth industrial revolution is driven by artificial intelligence (AI) and blockchain technology. It is bringing intelligence to production (Marwala and Xing, 2018). The social consequence of intelligent robots is extensive (Marwala, 2018; Xing and Marwala, 2018). Those with financial capital will simply buy these intelligent robots and produce goods and services to maximise profit. So the concept of the poor getting poorer and the rich getting richer will be exacerbated to the worse levels by the unconstrained drive towards intelligent automation. The Gini coefficient, which is a measure of inequality in society, will increase and this will threaten the very existence of the notion of the nation state. If people are going to be put out of their jobs who will buy these goods that these robots will produce? Where will these customers get the money to buy these goods if employment will not exist?

Research has shown that intelligent automation will decimate the number of jobs available due to the increasing use of automation from robots. Furthermore, the loss of jobs will reduce the amount of tax collected. So the primary consequences of automation is the decrease in tax revenue and employment. Bill Gates the founder of the giant computer company Microsoft proposes the taxation of AI robots to mitigate against the fall of tax revenue (French, 2018). South African-born technology billionaire Elon Musk proposes introducing the universal basic income to deal with the resulting loss of income due to the decrease in employment and personal income (Musk, 2018). If we are to assume that the proposed taxation of robots is a viable solution to counter the triple challenge of the decrease in employment, tax revenue and personal income, we ought to know what a robot is.

## 2.  What is a Robot?

The word robot comes from the Czech word "roborti" which means forced labor. According to the dictionary a robot is "any form of machinery that is able to perform a task or function automatically". Another definition from a dictionary defines a robot as "a machine that resembles a human and does mechanical, routine tasks on command". There are two classification of robots and these are intelligent and simple robots. Simple robots are those that execute exact commands. A vending machine, which is shown in Figure 1, replaces labor but is not intelligent because a customer comes and selects an option, pays, the choice is dispensed and the change is returned if any. This replaces the task of going to a shop and making a choice and exchanging money for goods from a shopkeeper. Perhaps, the intelligent aspect of a vending machine is the process of identifying money, calculating and giving back the change. The fact that vending machines put shopkeepers out of the work is certain. An example of an intelligent robot is shown in Figure 2. This robot can see, walk, perform tasks and has artificial intelligence to assist it to make good decisions. Conventionally, it is popular to postulate that automation came with artificial intelligence. For example, the assembly line of the second industrial revolution was automated to a certain extent. The traffic lights in our roads replaced a human traffic director and thus are examples of a robot. In South Africa, the traffic lights are even called robots.

When one wants to produce a document one can use a Microsoft (MS) Word. Microsoft Word is able to auto save documents, a feat which used to be performed by human beings. It is able to do automatically check the spelling. Furthermore, it is able to check the grammar. All these attributes used to be done by human beings. Companies used to employ people to perform these tasks. Does this mean that MS Word is a robot? If we were to make laws that tax robotic attributes, how do we tax this in addition to value added taxes?

The advent of artificial intelligence has introduced intelligence into robots and thus has accelerated the pace and complexity of automation. The next section studies the impact of the adoption of robots on taxation.

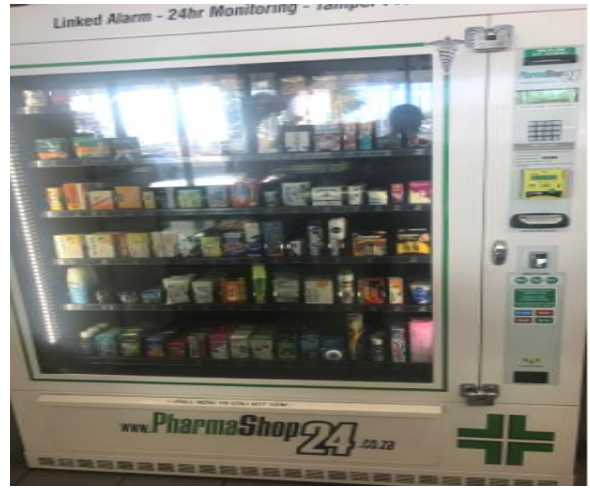

Figure 1. A picture of a vending machine

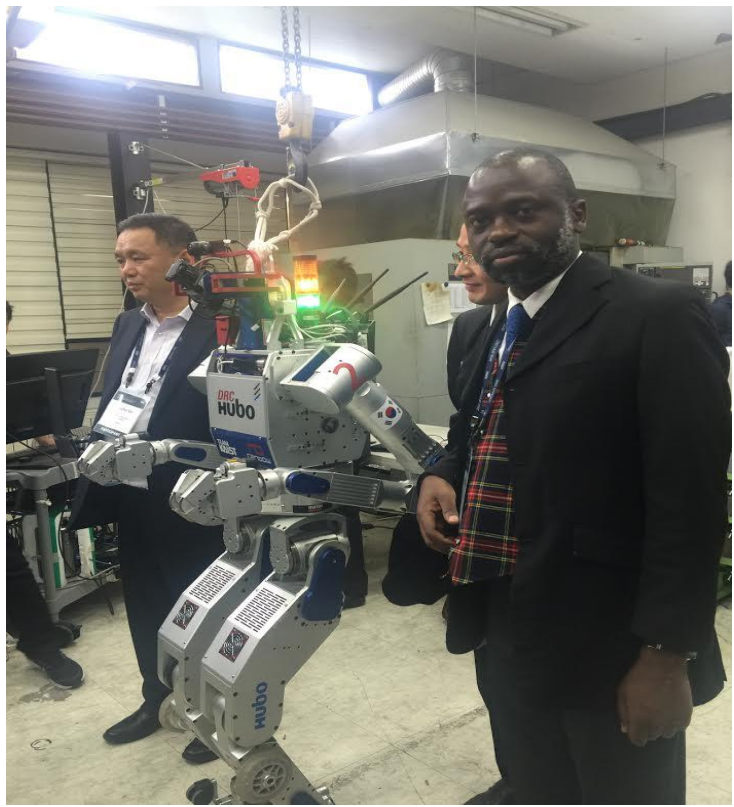

Figure 2. A picture of a humanoid robot from the Korean Advanced Institute of Science and Technology KAIST

## 3. Taxation

Tax collection is the most significant way in which government generates its revenue. Tax is used to run the state, to pay civil servants, to run public schools and hospitals to maintain roads as well as to pay social grants. These are public goods because they are society's investment into the stability and prosperity of a nation. If tax collection is higher than the expenditure then government runs a budget surplus. Alternatively, if tax collection is lower than the expenditure then government runs a budget deficit. A persisting deficit in a country ultimately renders a government bankrupt. The big question that needs to be answered is what will intelligent automation do to tax collection? Suppose the auto-manufacturing plant has 100 people who contribute $5000 per month as tax. If this factory decides to automate the jobs that these 100 employees are doing then the tax revenue for tax collection agency will drop by $500 000. The only way this tax revenue will not drop is if tax collection agency correspondingly increase corporate tax for this company by $500 000. Corporate tax is the tax that is charged to companies. The concept of increasing corporate tax because of automation is effectively taxing robots.

Can we realistically automate an entire economy? To answer this question we should study the work of Nobel Laureate Arthur Lewis. He proposed what later became known as the two sector economy theory. In this theory Lewis studies how to move an economy from agriculture to industry. He proposed that by moving labor from the agricultural sector to the industrial sector and using the resulting profits to expand industrial production can move a country from a developing to a developed economy. He proposed that this can be done until it is no longer economical to move labor from agricultural to industrial sector and this point is called the Lewis turning point. After the Lewis turning point has been reached the cost of labor in that economy starts rising. China used this theory for its developmental agenda and reached its Lewis turning point in 2009 after which the cost of labor started rising (Das and N'Diaye, 2013). Marwala and Hurwitz (2017) applied the Lewis' theory to automation and defined the limits of automation. They divided the economy into two parts and these were the AI machine and the human economy. Labor will move from the human economy to the AI machine economy until it no longer makes economic sense to migrate labor from humans to AI machines. This might be because the tasks involved are too complicated or too human to be automated. For example, if one needs to see a doctor perhaps this patient-doctor relationship is too human for people to prefer machine doctors over human doctors. So the limit of automation is a point at which it is more expensive to automate a job than to use humans mainly due to the difficulty of deploying technology.

Going back to tax collection, an American Economist Arthur Laffer in 1974 proposed a theory showing the relationship between the effective tax rate and the tax revenue collected by government (Zonana, 1985). These ideas popularized by Arthur Laffer were not new and were described as early as the 14$^{th}$ century by an Arab scholar Ibn Khaldun (Rabi, 1967). The Laffer curve, which is shown in Figure 3, states that if the government gives a tax rate of 0% it will collect no tax at all. If it gives a tax rate of 100%, people will have no incentive to work, and the taxes collected will be zero. Somewhere between the effective tax rate of 0% and 100%, there is a tax level that results in the maximum tax collected. Without automation this maximum tax collected will be a certain number. If we automate our factories without changing the corporate tax rate, then the tax collected will decrease. If we automate our factories and responsibly increase corporate tax rate, we can end up with the amount of tax collected which is higher than that collected without automation. This is shown in Figure 3.

Given all these, what do governments do to ensure that they collect enough taxes that will ensure a growing economy and consequently a stable society? Firstly, they should have dynamic and world class economic scientists who are sufficiently skilled in modelling economic, social and political phenomena. Secondly, they need to sufficiently trained people who understand automation and artificial intelligence. Thirdly, they should use AI robots to detect and prevent tax evasion. Fourthly, they should develop a framework to tax companies that are domiciled overseas but make their money in the country. These companies include Uber which connects taxi drivers to customers in the world from California and Netflix which runs Television in a country such as Spain from California.

Going back to the example of the MS Word, how should we tax this software given the fact that it contains automated elements? Should we ring fence those automated elements and tax them as we would have done if they were human beings? This will be cumbersome and is difficult. The alternative is to increase corporate tax to match the drop in tax revenue corresponding to automation. This should be done delicately by taking into account of the Laffer curve and ensuring that this increase does not result in the decrease of tax revenue due to the drop in incentive to corporates.

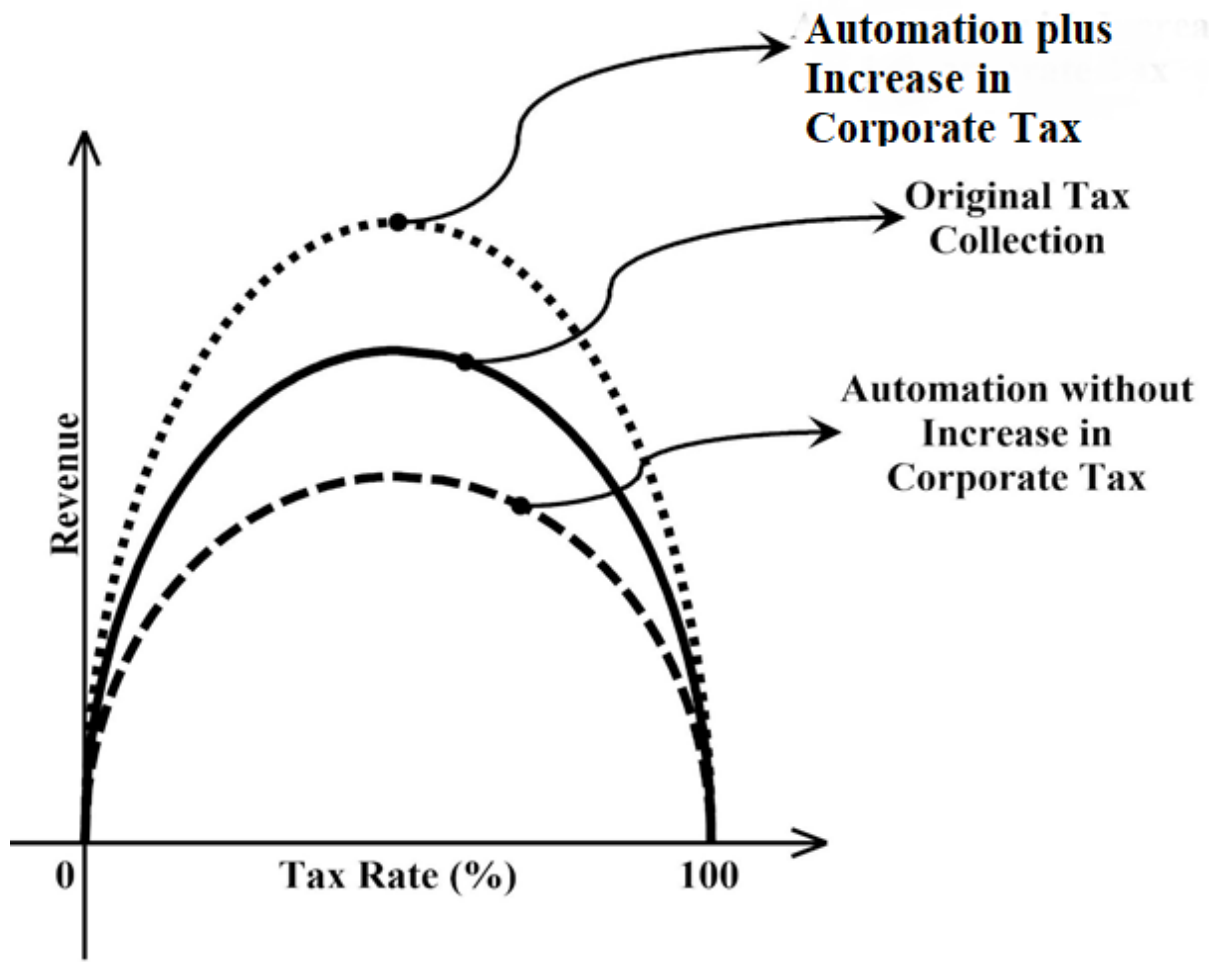

Figure 3 The Laffer Curve comparing automation with no automation

## 4. Conclusion

In conclusion, the fourth industrial revolution should not lead to the drop in the standards of living. Aristotle foretold this when he said: "The end of labor is to the gain of leisure". To ensure that indeed this is realized we should ensure that we manage the delicate balance amongst labor, automation and taxation. The drop in tax revenue should corresponded by the increase in corporate tax. It is evident from this article that it is a difficult task to establish what a robot is and what is not a robot.